\documentclass[final,5p,times,twocolumn]{elsarticle}

\usepackage[T1]{fontenc}
\usepackage[utf8]{inputenc}
\usepackage[swedish,english]{babel}
\biboptions{sort&compress}

\usepackage{amsmath,amssymb,xcolor,lineno,hyperref,array,bm,siunitx,booktabs}
\journal{Computer Physics Communications}

\definecolor{dr}{rgb}{.5,0,0}
\DeclareMathOperator{\re}{Re}
\DeclareMathOperator{\im}{Im}

\newcommand{\pt}{\partial}
\newcommand{\pe}{\perp}
\newcommand{\pa}{\parallel}
\newcommand{\be}{\begin{equation}}
\newcommand{\ee}{\end{equation}}
\def\ba#1\ea{\begin{align}#1\end{align}}
\newcommand{\br}{\left\{\begin{array}{*3{>{\displaystyle\vspace{1mm}}l}}}
\newcommand{\er}{\end{array}\right.}
\newcommand{\nn}{\nonumber}
\newcommand{\dd}[2]{\frac{\partial #1}{\partial #2}}
\newcommand{\dr}[2]{\frac{\mathrm{d} #1}{\mathrm{d} #2}}
\newcommand{\mc}{\mathcal}
\newcommand{\mr}{\mathrm}
\newcommand{\rme}{\mathrm{e}}
\newcommand{\rmi}{\mathrm{i}}
\newcommand{\rmd}{\mathrm{d}}
\newcommand{\rmc}{\mathrm{c}}
\newcommand{\rmB}{\mathrm{B}}
\newcommand{\rmp}{\mathrm{p}}
\newcommand{\rms}{\mathrm{s}}
\newcommand{\rmg}{\mathrm{g}}
\newcommand{\rmA}{\mathrm{A}}
\newcommand{\rmL}{\mathrm{L}}
\newcommand{\rmb}{\mathrm{b}}

\newcommand{\etal}{\emph{et al.}}

\begin{document}

\begin{frontmatter}



\title{FOXTAIL: Modeling the nonlinear interaction between\\Alfvén eigenmodes and energetic particles in tokamaks}


\author{\corref{a}Emmi Tholerus}
\ead{emmi.tholerus@ee.kth.se}

\author{\corref{}Thomas Johnson}
\ead{thomas.johnson@ee.kth.se}

\author{Torbjörn Hellsten}
\ead{torbjorn.hellsten@ee.kth.se}

\cortext[a]{Corresponding author}
\address{KTH Royal Institute of Technology, School of Electrical Engineering, Department of Fusion Plasma Physics, SE-100\,44 Stockholm, Sweden}

\begin{abstract}
FOXTAIL is a new hybrid magnetohydrodynamic--kinetic code used to describe interactions between energetic particles and Alfvén eigenmodes in tokamaks with realistic geometries. The code simulates the nonlinear dynamics of the amplitudes of individual eigenmodes and of a set of discrete markers in five-dimensional phase space representing the energetic particle distribution. Action--angle coordinates of the equilibrium system are used for efficient tracing of energetic particles, and the particle acceleration by the wave fields of the eigenmodes is Fourier decomposed in the same angles. The eigenmodes are described using temporally constant eigenfunctions with dynamic complex amplitudes. Possible applications of the code are presented, e.g., making a quantitative validity evaluation of the one-dimensional bump-on-tail approximation of the system. Expected effects of the fulfillment of the Chirikov criterion in two-mode scenarios have also been verified.
\end{abstract}


\end{frontmatter}

% \linenumbers

\section{Introduction}

Toroidal Alfvén eigenmodes (TAEs) are discrete frequency MHD waves that exist in the toroidicity induced gap of the Alfvén continuum spectra in toroidal magnetized plasmas. TAEs are typically excited by an ensemble of energetic ions (e.g.\ coming from auxiliary heating or from fusion reactions) with an inverted energy distribution along the characteristic curves of wave--particle interaction in momentum space. If these waves are excited to large amplitudes, they might eject a large fraction of energetic ions from the plasma before the ions transfer their energy to the bulk plasma, causing a significant reduction of heating efficiency of fast ions~\cite{hei,won}. It is therefore of great importance to understand the significance of TAEs in future devices, such as ITER. Accurate modeling is required that can resolve the nonlinear evolution of the wave--particle interactions.

Many hybrid MHD--kinetic Monte Carlo codes developed for this purpose are orbit following, requiring temporal resolutions well below bounce time scales of the resonant energetic particles. These are many orders of magnitude shorter than the time scales for the relevant dynamics of long-lived eigenmodes in large tokamaks, such as ITER. Relatively simple equations of motion that can resolve the relevant time scales for wave--particle interaction more efficiently can be acquired by using action--angle coordinates~\cite{kau} of the equilibrium system for the phase space of energetic particles. Particles in the unperturbed equilibrium system then follow straight lines in configuration space at constant velocities, and their canonical momenta are constants of motion. This coordinate system has the advantage that particles exactly remain on their guiding center orbits in the unperturbed system, independently of the time step length.

In order to get satisfactory convergence of the Monte Carlo codes describing nonlinear wave--particle interactions, $\delta f$ methods are often used. Although the method is computationally advantageous, it makes the code more difficult to use in conjunction with existing Monte Carlo codes that use full-$f$ methods or that use a different background distribution.

FOXTAIL (``\underline{FO}urier series e\underline{X}pansion of fas\underline{T} particle--\underline{A}lfvén eigenmode \underline{I}nteraction''-mode\underline{L}) is a new hybrid magnetohydrodynamic--kinetic model that both uses action--angle coordinates for particle phase space and a full-$f$ Monte Carlo method to represent the resonant energetic particle distribution. It is based on a model by Berk \etal~\cite{bb1}, which is derived from a Lagrangian formulation of the wave--particle interaction. The use of action--angle variables can give scenarios where the shortest time scale needed to be resolved in FOXTAIL is on the order of $\omega_\rmp^{-1}$, where $\omega_\rmp$ is the precession frequency of the energetic particles interacting strongly with the eigenmodes.

A simplification used in FOXTAIL is that the spatial structures of the eigenmode wave fields are taken to be constant in time. This limitation means that FOXTAIL is unable to model, e.g., energetic particle modes. There also exist scenarios where the time evolution of TAE eigenfunctions is of importance (see e.g.\ Ref.~\cite{wan}). Such scenarios can be modeled by other existing codes, not having the limitation of static eigenfunctions, such as the hybrid MHD--gyrokinetic code, HMGC~\cite{hm1,hm2,hm3}, and the gyrokinetic toroidal code, GTC~\cite{lin,zha,xia}.

The structure of this paper is the following: Section~\ref{sec:mod} presents the mathematical, the physical and the technical background of FOXTAIL, including derivations of the equations used in the different parts of the code and the overall structure of the code. Section~\ref{sec:coi} describes how some of the resulting equations are numerically implemented. Section~\ref{sec:num} presents some of the possible applications of FOXTAIL, including quantitative comparisons with the corresponding one-dimensional bump-on-tail approximation of the system and numerical studies of the Chirikov criterion in scenarios with two eigenmodes. Section~\ref{sec:con} summarizes the paper.

\section{Model description}\label{sec:mod}

\subsection{Physics overview}

The equations used in FOXTAIL are based on a Lagrangian formulation of the wave--particle system \cite{bb1}. Each simulation is formulated as an initial value problem, starting from an energetic particle distribution in a background equilibrium plasma and a set of eigenmodes with a static spatial structure and a dynamical amplitude and phase. Eigenmodes are treated as weak perturbations of the equilibrium, excluding direct mode--mode interaction. The nonlinear coupling of eigenmodes is taken into account only by the energy and momentum exchange between the modes via the energetic particles.

The physical process of the considered wave--particle interactions is essentially absorption and stimulated emission, and the interactions are energy and momentum conserving. The total energy of the eigenmode is proportional to the amplitude squared. Energy conservation is ensured by the equation
\be\label{eq:enc}
	\sum_\mr{particles}\dot{W}_\mr{wave} + \sum_\mr{eigenmodes}C\re(A \dot{A}^*) = 0,
\ee
where $\dot{W}_\mr{wave}$ is the time derivative of the kinetic particle energy, as accelerated by the wave field, $A$ is the complex amplitude of the eigenmode, and $C$ is the ratio between the eigenmode energy and $|A|^2/2$. A Lagrangian formulation of the wave--particle interaction is presented in section~\ref{sec:cse}, which is shown to be consistent with eq.~\eqref{eq:enc}.

The acceleration of a particle by an Alfvén eigenmode convects the momentum of the particles along curves in the phase space $(W, P_\phi, \mu)$ according to
\be\label{eq:cha}
	\frac{\rmd W}{\omega} = \frac{\rmd P_\phi}{n},~\rmd\mu = 0,
\ee
where $W$ is the particle kinetic energy, $P_\phi$ is the toroidal canonical momentum, $\mu$ is the magnetic moment, $n$ is the toroidal mode number of the eigenmode and $\omega$ is the eigenmode frequency. The magnetic moment is unperturbed, since the Alfvén eigenmodes are low frequency waves ($\omega \ll \omega_\rmc$). The curves in $W,P_\phi$-space specified by eq.~\eqref{eq:cha} are referred to as the characteristic curves of wave--particle interaction. Within certain parameter limits, FOXTAIL is equivalent with a one-dimensional bump-on-tail model describing the wave--particle interaction (cf.\ Ref.~\cite{bb2}). Different characteristic curves are indistinguishable in this limit, and the momentum of the 1D model then corresponds to a variable indexing the location of the particle along the characteristic curves. From the theory of the 1D bump-on-tail model, it is apparent that an inverted energy distribution along the characteristic curves at the wave--particle resonance can excite eigenmodes via a process analogous to Landau damping. 

\begin{figure}[t!]
\centering
\includegraphics[width=75mm]{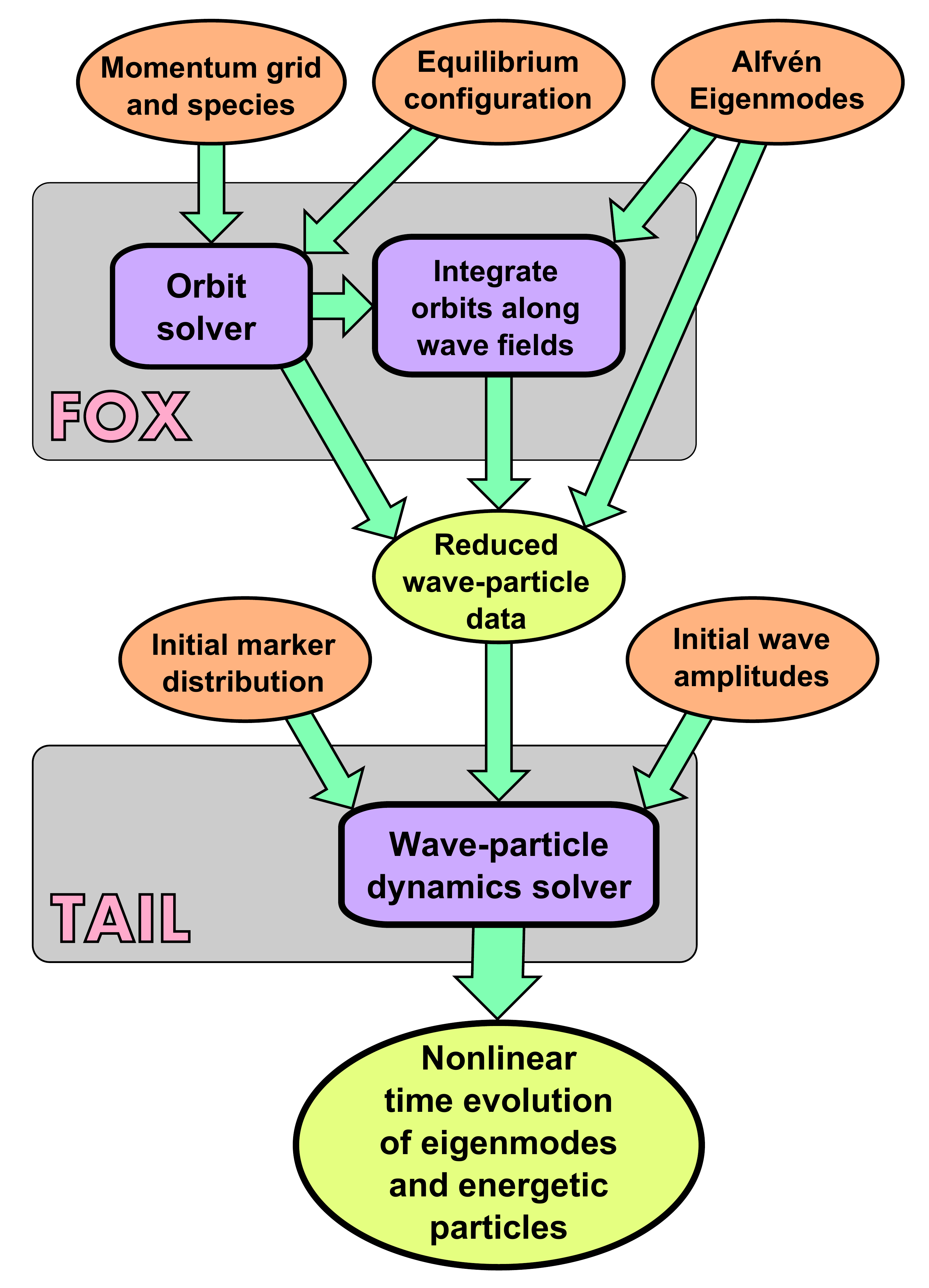}
\caption{Flowchart of the FOXTAIL code. As input, FOX takes the mass and charge of the energetic particle species and a grid in the space spanned by the adiabatic invariants of the equilibrium motion. On this grid, the orbits are solved using the input equilibrium configuration. External routines are used to calculate spatial wave field structures and frequencies of a chosen set of Alfvén eigenmodes. Eigenmode data and orbit data is sent to a routine that integrates all guiding center orbits along the wave fields of the eigenmodes in order to obtain a set of interaction coefficients characterizing the particle response with respect to the wave. A set of particle, eigenmode and wave--particle interaction data is collected and sent to the TAIL code, along with an initial distribution of markers and initial complex amplitudes of the eigenmodes. TAIL then solves the nonlinear time evolution of the markers and the considered eigenmodes.}\label{fig:fch}
\end{figure}

\subsection{Overview of the FOXTAIL code}

The FOXTAIL code is essentially split into two parts, ``FOX'' and ``TAIL'', as illustrated in Fig.~\ref{fig:fch}. TAIL is the numerical dynamics solver of the eigenmodes and the energetic particles, solving the wave--particle system as an initial value problem. FOX can be viewed as a preprocessor of TAIL, calculating orbital, eigenmode and interaction related data for a given set of eigenmodes and energetic particle species in a defined equilibrium configuration. The distribution of energetic particles is represented using a finite set of markers with predefined weights.

\subsection{Action--angle coordinates}\label{sec:aac}

The phase space of markers in TAIL is described using action--angle coordinates of the equilibrium system \cite{kau}. The equations of motion of the system become particularly simple in this choice of coordinates, which contributes to the computational efficiency of the solver. In the absence of wave field perturbations, the ``action'' coordinates (i.e.\ the momentum coordinates of the canonical action--angle coordinate system) are constants of motion, whereas the ``angle'' coordinates (configuration space coordinates) evolve with constant velocities. The presence of eigenmodes perturbs this simple dynamics, and the adiabatic invariants are convected according to eq.~\eqref{eq:cha}.

The angle coordinates of the system are given by
\be\label{eq:ang}\br
	\tilde{\alpha} = \alpha + \int_0^\theta\rmd\theta'\frac{\bar{\omega}_\rmc(\bm{P}) - \omega_\rmc(\bm{P}, \theta')}{\dot{\theta}(\bm{P}, \theta')}, \\
	\tilde{\theta} = \int_0^\theta\rmd\theta'\frac{\omega_\rmB(\bm{P})}{\dot{\theta}(\bm{P}, \theta')}, \\
	\tilde{\phi} = \phi + \int_0^{\theta}\rmd\theta'\frac{\omega_\rmp(\bm{P}) - \dot{\phi}(\bm{P}, \theta')}{\dot{\theta}(\bm{P}, \theta')},
\er\ee
where $\alpha$ is the gyro-angle, $\theta$ and $\phi$ are the poloidal and toroidal angles, respectively, $\bm{P} \equiv (P_\alpha, P_\theta, P_\phi)$ are the action coordinates, canonical to the angles $(\tilde{\alpha}, \tilde{\theta}, \tilde{\phi})$. Furthermore, $\omega_\rmc$ ($\bar{\omega}_\rmc$) is the (time averaged) gyro-frequency and $\omega_\rmB$ and $\omega_\rmp$ are the bounce and precession frequencies, respectively. The integrals of eq.~\eqref{eq:ang} are evaluated on the poloidal coordinates along the guiding center orbits. 

FOXTAIL uses the adiabatic invariants $\bm{J} \equiv (\mu, \Lambda, P_\phi)$ as momentum coordinates ($\Lambda = \mu B_0/W$ is the normalized magnetic moment, where $B_0$ is the on-axis magnetic field strength). These momentum coordinates can be expressed as functions of $\bm{P}$. The angle coordinates $(\tilde{\alpha}, \tilde{\theta}, \tilde{\phi})$ are referred to as the \emph{transformed} gyro-angle, poloidal angle and toroidal angle, respectively. In the equilibrium system, where the momentum coordinates $\bm{J}$ are constant, the transformed angles evolve at a constant velocity $(\bar{\omega}_\rmc, \omega_\rmB, \omega_\rmp)$. The dynamics in the FOXTAIL model is averaged over gyration time scales, and consequently $\tilde{\alpha}$ is an ignorable coordinate of the system.

The transformed poloidal angle, $\tilde{\theta}$, can be viewed as an index of the location of the particle in the guiding center orbit, where $\tilde{\theta}: 0 \to 2\pi$ is a complete period of the guiding center orbit in the poloidal plane ($\tilde{\theta} = 0$ is defined as the point where the outer leg of the orbit intersects with the equatorial plane). In section~\ref{sec:wav}, it is shown that a Fourier expansion of the instant acceleration of the particle in the wave field is a convenient representation of the wave--particle interactions.

Complications arise for the $\tilde{\theta}$ coordinate close to the boundary $\omega_\rmB = 0$, where the particle asymptotically approaches one of the turning points. In this limit, all points along the orbit besides the turning points are represented by infinitely narrow intervals in $\tilde{\theta}$, and the representation of the wave--particle interaction using Fourier series expansions in $\tilde{\theta}$-space becomes invalid. These complications can potentially be resolved either by ad hoc boundary conditions or by more sophisticated coordinate transformations close to this boundary. However, all of the numerical studies presented in this paper consider scenarios where the simulated energetic particle distributions are sufficiently far from the boundary.

\subsection{Orbit solver}

FOX contains subroutines that solves the 3D motion of particles in a given equilibrium field configuration on a grid in $\bm{J}$-space. The equilibrium configuration and the particle orbits are described in $\psi,\theta,\phi$-space, where $\psi$ is the poloidal magnetic flux per radian. For each orbit on the $\bm{J}$-grid, the time evolution of $\psi$, $\theta$ and $\phi_\rms$ is calculated, where $\phi_\rms \equiv \phi - \tilde{\phi} + \omega_\rmp\tilde{\theta}/\omega_\rmB$ is the shifted toroidal coordinate ($\phi_\rms$ is $\phi$ shifted such that $\phi_\rms = 0$ coincides with $\tilde{\theta} = 0$). The equilibrium configuration is taken to be axisymmetric. Assuming MHD force balance and nested magnetic flux surfaces, the equilibrium magnetic field can be expressed as
\be
	\bm{B} = F(\psi)\nabla\phi + \nabla\phi\times\nabla\psi.
\ee

The guiding center motion consists of a parallel motion and a drift motion, where the drift is given by the combined $\bm{E}\times\bm{B}$, $\nabla B$ and curvature drifts according to
\be
	\bm{v}_\rmd = \frac{\bm{E}\times\bm{B}}{B^2} + \frac{\mu(2 B_0 - \Lambda B)}{Z e \Lambda B^3}\bm{B}\times\nabla B.
\ee
By combining $W = m v_\pa^2/2 + \mu B$ with $P_\phi = m v_\pa F/B + m v_{\rmd,\phi} - Z e\psi$ and eliminating $v_\pa$, it can be shown that the coordinates of the guiding center orbits in the poloidal plane follow the equation 
\be\label{eq:gco}
	f(\psi, \theta, \bm{J}) = 0,
\ee
where
\ba
	f(\psi, \theta, \bm{J}) &\equiv 1 - \frac{\Lambda B(\psi, \theta)}{B_0} - \frac{\Lambda B^2(\psi, \theta)}{2 m \mu B_0 F^2(\psi)} \nn\\
	&\quad\, \times [P_\phi + Z e\psi - m v_{\rmd, \phi}(\psi, \theta, \mu, \Lambda)]^2, 
\ea
\ba
	v_{\rmd, \phi}(\psi, \theta, \mu, \Lambda) &= \frac{E^\psi(\psi, \theta)}{B^2(\psi, \theta)} - \frac{\mu[2 B_0 - \Lambda B(\psi, \theta)]}{Z e \Lambda B^3(\psi, \theta)}\nn\\
	&\quad\, \times[\nabla B(\psi, \theta)]^\psi
\ea
is the covariant toroidal component of the drift velocity, $E^\psi$ and $[\nabla B]^\psi$ are the contravariant $\psi$-components of $\bm{E}$ and $\nabla B$, respectively, and $m$ is the particle mass.

Once the projections of the orbits on the poloidal plane are solved on the chosen $\bm{J}$-grid using eq.~\eqref{eq:gco}, the corresponding time coordinates are calculated according to
\be\label{eq:tps}
	t(\psi) = \int_{\psi_0}^\psi\frac{\rmd\psi}{\dot{\psi}},
\ee
where
\be\label{eq:dps}
	\dot{\psi} = \frac{J F E_\theta - g_{\theta\theta}E_\phi}{J^2 B^2} - \frac{\mu F(2 B_0 - \Lambda B)}{Z e \Lambda J B^3}\dd{B}{\theta},
\ee\be
	J \equiv \left(\dd{\bm{r}}{\psi}\times\dd{\bm{r}}{\theta}\right)\cdot\dd{\bm{r}}{\phi},
\ee
and $\psi(t = 0) = \psi_0$ is defined as the point where the outer leg of the guiding center orbit intersects the equatorial plane ($\tilde{\theta} = 0$). Similarly, the $\phi_\rms$ coordinates are calculated from the toroidal velocity according to
\be\label{eq:tph}
	\phi_\rms(t) = \int_0^t\rmd t'\:\dot{\phi}(\bm{J}, t'),
\ee
where
\be\label{eq:dph}
	\dot{\phi}(\bm{J}, t) = \frac{P_\phi + Z e\psi(\bm{J}, t)}{m R^2(\bm{J}, t)}.
\ee
When calculating the wave--particle interaction coefficients, the poloidal velocity is required at each point of the orbit (see eq.~\eqref{eq:vig}). It is given by
\ba\label{eq:dth}
	\dot{\theta} &= \frac{P_\phi + Z e\psi - m v_{\rmd,\phi}}{m J F} - \frac{J F E_\psi + g_{\psi\theta}E_\phi}{J^2 B^2} \nn\\
	&\quad\, + \frac{\mu F(2 B_0 - \Lambda B)}{Z e \Lambda J B^3}\dd{B}{\psi}.
\ea

\subsection{Wave field description}\label{sec:wav}

The present version of FOXTAIL describes the dynamics of low frequency, \emph{shear} eigenmodes, such as the toroidal Alfvén eigenmodes (TAEs), but the model can be extended to describe eigenmodes with, e.g., compressible components as well. 

Neglecting plasma resistivity, the general electric wave field can be represented by the two scalar potentials $\Phi$ and $\Psi$ according to
\be\label{eq:esc}
	\delta\bm{E} = -\nabla_\pe\Phi + \frac{\bm{B}\times\nabla\Psi}{B},
\ee
where the first term is associated with magnetic shear, and the second term is associated with magnetic compression. When parallel gradients in the plasma are negligible in comparison to perpendicular gradients, the excitation of the two scalar potentials $\Phi$ and $\Psi$ is almost decoupled, and it is sufficient to describe the shear Alfvén wave using the first term \cite{bb1}.

In an axisymmetric toroidal plasma, the scalar potential of each eigenmode (indexed by $i$) can be written on the form
\ba
	\Phi_i(\bm{r}, t) &= \re\sum_m C_i(t)\rme^{\rmi\chi_i(t)}\Phi_{i, m}(\psi) \nn\\
	&\quad\, \times\rme^{\rmi(n_i \phi - m\theta - \omega_i t)}, \label{eq:phi}
\ea
where $C_i$ and $\chi_i$ are the slowly varying amplitude and phase of the eigenmode, respectively ($\dot{C}_i/C_i \ll \omega_i$, $\dot{\chi_i} \ll \omega_i$), $n_i$ is the toroidal mode number and $\omega_i$ is the eigenmode frequency. The electric wave field therefore given by
\be
	\delta\bm{E} = \re\sum_i C_i\rme^{\rmi\chi_i}\tilde{\bm{E}}_i\rme^{\rmi(n_i\phi - \omega_i t)},
\ee
where
\ba
	\tilde{\bm{E}}_i &= \sum_m\rme^{-\rmi m\theta}\bigg(\left[\rmi\Phi_{i,m}g_{\psi\theta}G_{i,m} - \dr{\Phi_{i, m}}{\psi}\right]\nabla\psi + \rmi\Phi_{i,m} \nn\\
	&\quad\,\times\Big[(g_{\theta\theta}G_{i,m} + m)\nabla\theta + (J F G_{i,m} - n_i)\nabla\phi\Big]\bigg), \label{eq:tie}
\ea
$G_{i,m} \equiv (n_i J F - m R^2)/(J^2 R^2 B^2)$.

\subsection{Fourier series expansion of fast particle--Alfvén eigenmode interaction}

The acceleration of the energetic particle in the wave field is described by the equation
\be
	\dot{W} = Z e \bm{v}\cdot\delta\bm{E} \approx Z e\langle\bm{v}\cdot\delta\bm{E}\rangle_\rmg,
\ee
where $\langle\cdot\rangle_\rmg$ averages over the gyro-motion. Initial versions of FOXTAIL only consider the lowest order averaging over the gyro-motion, but gyro-kinetic corrections to the averaging may be included in later versions. The averaged $\bm{v}\cdot\delta\bm{E}$ can be written as
\be
	\langle\bm{v}\cdot\delta\bm{E}\rangle_\rmg = \langle\bm{v}\rangle_\rmg\cdot\langle\delta\bm{E}\rangle_\rmg + \frac{\mu}{Z e}\dd{\langle\delta B_\pa\rangle_{S_\rmg}}{t},
\ee
where $\langle\delta B_\pa\rangle_{S_\rmg}$ averages the parallel magnetic wave field over the surface $S_\rmg$ enclosed by the gyro-ring (generated by the span of the gyro-angle. For shear waves, the $\delta B_\pa$ term can be neglected, and we are left with
\be
	\dot{W} = Z e\langle\bm{v}\rangle_\rmg\cdot\langle\delta\bm{E}\rangle_\rmg = \re\sum_i C_i\rme^{\rmi\chi_i} V_i \rme^{\rmi(n_i\phi - \omega_i t)},
\ee
where
\be\label{eq:vig}
	V_i(\bm{J}, t) = Z e(\dot{\psi}\tilde{E}_\psi + \dot{\theta}\tilde{E}_\theta + \dot{\phi}\tilde{E}_\phi),
\ee
$\dot{\psi}(\bm{J}, t)$, $\dot{\theta}(\bm{J}, t)$ and $\dot{\phi}(\bm{J}, t)$ are the guiding center velocities, given by eqs.~\eqref{eq:dps}, \eqref{eq:dth} and \eqref{eq:dph}, respectively, and $\tilde{E}_\psi(\psi(t), \theta(t))$, $\tilde{E}_\theta(\psi(t), \theta(t))$ and $\tilde{E}_\phi(\psi(t), \theta(t))$ are given by eq.~\eqref{eq:tie}.

All guiding center coordinates $(\psi, \theta, \phi_\rms)$ and their velocities can be written as functions of $\bm{J}$ and $\tilde{\theta}$. It is then possible to write $\dot{W}$ on a very compact form using action--angle coordinates:
\be\label{eq:dwi}
	\dot{W} = \re\sum_{i,\ell}C_i\rme^{\rmi\chi_i}V_{i,\ell}\rme^{\rmi(\ell\tilde{\theta} + n_i\tilde{\phi} - \omega_i t)},
\ee
where
\ba
	V_{i,\ell}(\bm{J}) &= \frac{Z e}{2\pi}\int_0^{2\pi}\rmd\tilde{\theta}\:V_i(\tilde{\theta}, \bm{J}) \nn\\
	&\quad\, \times\exp\bigg[\rmi\bigg(n_i\left[\phi_\rms(\tilde{\theta}, \bm{J}) - \frac{\omega_\rmp(\bm{J})}{\omega_\rmB(\bm{J})}\tilde{\theta}\right] - \ell\tilde{\theta}\bigg)\bigg]\label{eq:vil}
\ea
The coefficients $V_{i, \ell}$ are the Fourier series expansion of the wave--particle interaction that named the FOX code. It can be seen in eq.~\eqref{eq:dwi} that a wave--particle resonance is characterized by the condition $\ell\omega_\rmB + n_i\omega_\rmp \approx \omega_i$. The model is very efficient in the sense that one can select the relevant modes $i$ and coefficients $\ell$ that are close to resonant with an ensemble of energetic particles and neglect all other modes and non-resonant Fourier coefficients. If the relevant coefficients only have $\ell = 0$, $\tilde{\theta}$ becomes an ignorable coordinate of the system, and the shortest time scales that need to be resolved in simulations using TAIL are the precession time scales, $\omega_\rmp^{-1}$.

As was mentioned in section~\ref{sec:aac}, complications arise for the choice of action--angle coordinates when describing the wave--particle interaction in regions where $\omega_\rmB \approx 0$. This is explicitly seen in eq.~\eqref{eq:vil}, where $\omega_\rmp/\omega_\rmB \to \infty$, and the integrand oscillates infinitely fast in $\tilde{\theta}$. $V_i(\tilde{\theta}, \bm{J})$ is bounded, and constant for all $\tilde{\theta}$ in this limit except for the infinitely narrow intervals that do not represent the turning points. For this reason, it can be understood that $V_{i,\ell}$ tends to zero towards the $\omega_\rmB = 0$ boundary surface. Note that the sum of eq.~\eqref{eq:dwi} over all integers $\ell \in \mathbb{Z}$ should remain finite, given that particles can be accelerated across these surfaces by wave fields.

\subsection{Lagrangian formulation of wave--particle interaction}\label{sec:cse}

A model for describing the dynamics of the momentum variables $(\mu, \Lambda, P_\phi)$ and the amplitudes and phases of the eigenmodes remains to be found. Such a model can be derived from a Lagrangian formulation of the wave--particle system. We consider additions to the equilibrium Lagrangian by power series of the eigenmode amplitude $C_i$. 

Expressed in action--angle variables, the zeroth order Lagrangian reads \cite{whi}
\be\label{eq:la0}
	\mc{L}_0 = \sum_k \left[P_{\alpha,k}\dot{\tilde{\alpha}}_k + P_{\theta,k}\dot{\tilde{\theta}}_k + P_{\phi, k}\dot{\tilde{\phi}}_k - \mc{H}_0(\bm{P}_k)\right],
\ee
where $k$ is the particle index, and $\mc{H}_0$, satisfying
\be
	\dd{\mc{H}_0}{\bm{P}} = (\bar{\omega}_\rmc, \omega_\rmB, \omega_\rmp),
\ee
is the equilibrium Hamiltonian. A first order Lagrangian consistent with eq.~\eqref{eq:dwi} is
\be\label{eq:la1}
	\mc{L}_1 = \im\sum_{i,k,\ell}\frac{C_i}{\omega_i}\rme^{\rmi\chi_i}V_{i,\ell}(\bm{J}_k)\rme^{\rmi(\ell\tilde{\theta}_k + n_i\tilde{\phi}_k - \omega_i t)}.
\ee
In a low-$\beta$ plasma, the second order Lagrangian for shear Alfvén eigenmodes can be expressed on the form \cite{bb1}
\be\label{eq:l20}
	\mc{L}_2 = -\sum_i\frac{\dot{\chi}_i C_i^2}{2\mu_0\omega_i}\int\rmd V\:\frac{|\tilde{\bm{E}}_i(\bm{r})|^2}{v_\rmA^2(\bm{r})}
\ee
when neglecting rapidly oscillating terms (on time scales $\omega_i^{-1}$), where $\mu_0$ is the vacuum permeability. 

Note that the system is invariant under the transformation $C_i\rme^{\rmi\chi_i} \to \kappa_i C_i\rme^{\rmi\chi_i}$, $\Phi_{i,m} \to \Phi_{i,m}/\kappa_i$ for any constant $\kappa_i$. A normalization can be imposed by the condition
\be
	\int\rmd V\:\frac{|\tilde{\bm{E}}_i(\bm{r})|^2}{\mu_0 v_\rmA^2(\bm{r})} = 1
\ee
for each mode $i$. Then the second order Lagrangian reduces to
\be\label{eq:la2}
	\mc{L}_2 = -\sum_i\frac{\dot{\chi}_i C_i^2}{2\omega_i}.
\ee

Assuming that the amplitudes $C_i$ are small enough, such that $|\mc{L}_1| \ll |\mc{H}_0|$, the corresponding Hamiltonian for the wave--particle system is $\mc{H} = \mc{H}_0 - \mc{L}_1$, with the canonically conjugate pairs $(\tilde{\alpha}_k\,; P_{\alpha, k})$, $(\tilde{\theta}_k\,; P_{\theta, k})$, $(\tilde{\phi}_k\,; P_{\phi, k})$ and $(-\chi_i\,; C_i^2/2\omega_i)$. From this, one can derive all the remaining equations of motion of the wave--particle system self-consistently.

Defining the complex amplitude $A_i \equiv C_i\rme^{\rmi\chi_i}$ and using that $\mu = Z e P_\alpha / m$, the equations of motion of the wave--particle system is
\be\label{eq:ta1}\br
	\dot{\mu}_k = 0, & \dot{\tilde{\theta}}_k = \omega_\rmB(\bm{J}_k), \\
	\dot{\Lambda}_k = -\frac{\Lambda_k^2}{\mu_k B_0}\re\sum_i A_i U_{i,k}, & \dot{\tilde{\phi}}_k = \omega_\rmp(\bm{J}_k), \\
	\dot{P}_{\phi,k} = \re\sum_i\frac{n_i}{\omega_i}A_i U_{i,k}, & \dot{A}_i = -\sum_k U_{i,k}^*,
\er\ee
where
\be
	U_{i,k} \equiv \sum_\ell V_{i,\ell}(\bm{J}_k)\rme^{\rmi(\ell\tilde{\theta}_k + n_i\tilde{\phi}_k - \omega_i t)}.
\ee

\subsection{Additional operators}\label{sec:col}

The effects of particle collisions are not covered by the theory presented in the preceding sections. These effects can be added explicitly to the system, e.g.\ by using a diffusion operator in momentum space \cite{bb3} combined with a momentum drag \cite{lil} derived from the Fokker--Planck operator, which act directly on the energetic particle distribution. Adding diffusion to the system transforms the system of ordinary differential equations in eq.~\eqref{eq:ta1} to a system of stochastic differential equations. The general drag--diffusion operator can be written on It\=o form according to
\be\label{eq:col}\br
	\rmd\mu_\rmc = a_\mu\rmd t + \bm{b}_\mu\cdot\rmd\bm{W}_t, \\
	\rmd\Lambda_\rmc = a_\Lambda\rmd t + \bm{b}_\Lambda\cdot\rmd\bm{W}_t, \\
	\rmd P_{\phi,\rmc} = a_{P_\phi}\rmd t + \bm{b}_{P_\phi}\cdot\rmd\bm{W}_t,
\er\ee
where $a_{\bm{J}}$ is associated with drag, $\bm{b}_{\bm{J}}$ is associated with diffusion and the components of $\bm{W}_t$ are independent Wiener processes in time. Both $a_{\bm{J}}$ and $\bm{b}_{\bm{J}}$ are functions of $\bm{J}$ and $\tilde{\theta}$ in general, but orbit averaged (i.e.\ $\tilde{\theta}$ averaged) versions of the operators may be used for simplicity (see e.g.\ Ref.~\cite{eri}).

There are other processes external to the energetic particle--Alfvén eigenmode system which may be included. Depending on the source of the energetic particle distribution (e.g.\ neutral beam injection, cyclotron resonance heating or nuclear fusion), operators can be added that continuously supply particles and/or energy to the system. Particle sources are typically modeled by dynamical weights and statistical redistribution of markers. Losses from Bremsstrahlung and cyclotron radiation can be modeled by additions to the $a_{\bm{J}}$ terms in eq.~\eqref{eq:col}.

An additional $-\gamma_{\rmd,i} A_i$ term can be added to the amplitude equation ($\dot{A}_i$ in eq.~\eqref{eq:ta1}) in order to model various damping mechanisms on the eigenmode amplitudes, where $\gamma_{\rmd, i}$ is the damping rate of the $i$:th mode. This damping can, e.g., come from Landau damping in the interaction with thermal particles or damping due to mode conversion. The $-\gamma_{\rmd,i} A_i$ term of the amplitude equation is here referred to as \emph{explicit} wave damping, unlike the Landau damping coming from the interaction with the energetic particle distribution, which arises implicitly from the model equations.

% In the presence of the $\rmd\bm{W}_t$ terms in eq.~\eqref{eq:col}, the added angular perturbations $\pt\tilde{\theta}/\pt\bm{J}\cdot\rmd\bm{J}$ and $\pt\tilde{\phi}/\pt\bm{J}\cdot\rmd\bm{J}$ act as phase decorrelation operators on the systems, which were analyzed in Ref.~\cite{tho}.

\subsection{Lowest order corrections to the angle perturbation}\label{sec:loc}

When going from the Lagrangians of eqs.~\eqref{eq:la0}, \eqref{eq:la1} and \eqref{eq:la2} to eq.~\eqref{eq:ta1}, it was assumed that $|\mc{L}_1| \ll |\mc{H}_0|$, which allowed one to neglect corrections of the canonical momenta coming from the implicit dependence of $\mc{L}_1$ with respect to $\dot{\tilde{\theta}}$ and $\dot{\tilde{\phi}}$. For a small enough $\mc{L}_1$, the final equations for $\dot{\tilde{\theta}}$ and $\dot{\tilde{\phi}}$ are given only by the derivatives of the equilibrium Hamiltonian $\mc{H}_0$ with respect to $P_\theta$ and $P_\phi$, respectively. However, the above assumptions might not be valid for large amplitude eigenmodes and for processes affecting the energetic particles or the eigenmodes that are not direct wave--particle interaction, and one may have to consider the lowest order correction to $\dot{\tilde{\theta}}$ and $\dot{\tilde{\phi}}$ due to momentum perturbation. The correction can be understood from the fact that $\tilde{\theta}$ maps differently to locations on the guiding center orbit for different $\bm{J}$. Without the correction, one pushes the particle forwards or backwards along the orbit due to different mapping of $\tilde{\theta}$.

Assuming a minimal perturbation of the guiding center position in $R,Z$-space while perturbing $\bm{J}$ by the amount $\rmd\bm{J}$, it can be shown that $\tilde{\theta}$ should be corrected by the amount $\pt\tilde{\theta}/\pt\bm{J}\cdot\rmd\bm{J}$, where
\be\label{eq:dtj}
	\dd{\tilde{\theta}}{\bm{J}} = -\frac{\omega_\rmB}{\dot{R}^2 + \dot{Z}^2}\left(\dot{R}\dd{R}{\bm{J}} + \dot{Z}\dd{Z}{\bm{J}}\right),
\ee
$R$ is the distance from the symmetry axis and $Z$ is the vertical guiding center position. Using that $\tilde{\phi} = \phi - \phi_\rms + \omega_\rmp\tilde{\theta}/\omega_\rmB$, it can be shown that
\ba\label{eq:dpj}
	\dd{\tilde{\phi}}{\bm{J}} &= \frac{\dot{\phi} - \omega_\rmp}{\dot{R}^2 + \dot{Z}^2}\left(\dot{R}\dd{R}{\bm{J}} + \dot{Z}\dd{Z}{\bm{J}}\right). \nn\\
	&\quad\, + \dd{}{\bm{J}}\left(\frac{\omega_\rmp}{\omega_\rmB}\tilde{\theta} - \phi_\rms\right).
\ea
All the derivatives with respect to $\bm{J}$ on the right hand sides of eqs.~\eqref{eq:dtj} and \eqref{eq:dpj} are evaluated while keeping $\tilde{\theta}$ constant.

When the $\bm{J}$ perturbations of any process is stochastic ($\bm{b}_{\bm{J}}$ of eq.~\eqref{eq:col} is nonzero), the angle coordinates become stochastic as well when including the $\pt\tilde{\theta}/\pt\bm{J}$ and $\pt\tilde{\phi}/\pt\bm{J}$ corrections. This is typically how phase decorrelation is introduced to the system. Such an operator was analyzed for the one-dimensional bump-on-tail model in Ref.~\cite{tho}.

\subsection{Summary of the TAIL model equations}\label{sec:sum}

To summarize, the model equations used by TAIL are
\be\label{eq:ta2}\br
	\rmd\tilde{\theta}_k = \omega_\rmB(\bm{J}_k)\rmd t + \dd{\tilde{\theta}(\tilde{\theta}_k, \bm{J}_k)}{\bm{J}}\cdot\rmd\bm{J}_k, \\
	\rmd\tilde{\phi}_k = \omega_\rmp(\bm{J}_k)\rmd t + \dd{\tilde{\phi}(\tilde{\theta}_k, \bm{J}_k)}{\bm{J}}\cdot\rmd\bm{J}_k, \\
	\rmd \mu_k = \rmd\mu_\rmc(\bm{J}_k), \\
	\rmd\Lambda_k = -\frac{\Lambda_k^2}{\mu_k B_0}\re\sum_i A_i U_{i,k}\rmd t + \rmd\Lambda_\rmc(\bm{J}_k), \\
	\dot{P}_{\phi,k} = \re\sum_i\frac{n_i}{\omega_i}A_i U_{i,k} + \rmd P_{\phi,\rmc}(\bm{J}_k), \\
	\dot{A}_i = -\sum_k w_k U_{i,k}^* - \gamma_{\rmd, i} A_i,
\er\ee
where $k$ is now the \emph{marker} index with weight $w_k$, and $(\rmd\mu_\rmc, \rmd\Lambda_\rmc, \rmd P_{\phi,\rmc})$ represents additional differential operators acting on the momentum space of markers. The total wave--particle energy of the system can be defined as
\ba
	W_\mr{tot} &\equiv \sum_k w_k W_k + \sum_i\frac{|A_i|^2}{2} \nn\\
	&= \sum_k w_k \frac{\mu_k B_0}{\Lambda_k} + \sum_i\frac{|A_i|^2}{2}.
\ea
In the absence of explicit wave damping and particle sources and sinks, it can easily be shown that
\be
	\dot{W}_\mr{tot} = -\sum_k\frac{w_k \mu_k B_0}{\Lambda_k^2}\dot{\Lambda}_k + \re\sum_i A_i\dot{A}_i^* = 0,
\ee
which is consistent with the condition for energy conservation in eq.~\eqref{eq:enc}.

\section{Code implementation}\label{sec:coi}

\begin{figure}[t]
\centering
\includegraphics[width=58mm]{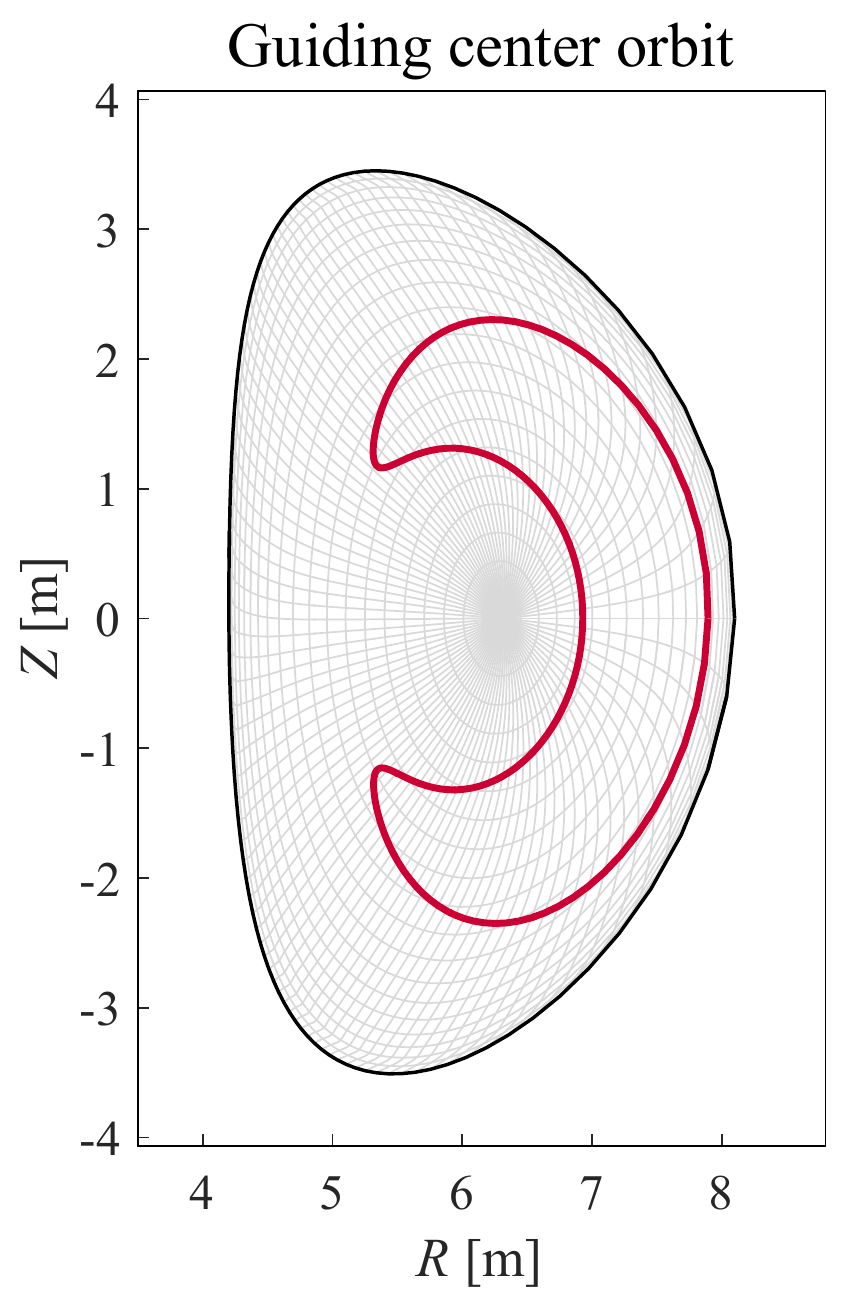}
\caption{Guiding center orbit in the poloidal plane solved by FOX in an ITER equilibrium, using a deuterium ion with $\mu = \SI{4}{MeV/T}$, $\Lambda = 0.85$ and $P_\phi = \SI{5}{eVs}$ (\SI{1}{eV} =  \SI{1.6022e-19}{J}).}\label{fig:orb}
\end{figure}

As input, FOX takes a file that contains all information characterizing the equilibrium configuration on a format compatible with Integrated Tokamak Modelling standards \cite{imb}. All scalar fields, such as $B$, $J$ and $F$, are specified on a grid in $\psi,\theta$-space. The user defines an equidistant grid in $\bm{J}$-space, where all guiding center orbits are to be solved in the given equilibrium. Equation~\eqref{eq:gco} is then solved for each $\bm{J}$ on the grid by bilinear interpolation of $f$ in $\psi,\theta$-space. An example of such a solution is shown in Fig.~\ref{fig:orb}. The solution method is optimal for wide orbits, whereas thinner orbits require a large enough $\psi$ resolution of the equilibrium.

Once the guiding center orbit points are found in the poloidal plane, the time dependence of the orbit is calculated using eq.~\eqref{eq:tps}. Numerically, the integration is performed by assuming $\ddot{\psi}$ to be constant between adjacent points of the guiding center orbit. This assumption generates the approximation
\be\label{eq:tpa}
	t_j = \int_{\psi_0}^{\psi_j}\frac{\rmd\psi}{\dot{\psi}} \approx 2\sum_{k = 1}^j\frac{\psi_k - \psi_{k - 1}}{\dot{\psi}_k + \dot{\psi}_{k - 1}},
\ee
where $(\psi_j, \theta_j)$ is the $j$:th $\psi,\theta$-point along the orbit and $\dot{\psi}_j$ is $\dot{\psi}$ evaluated at $(\psi_j, \theta_j)$. Equation~\eqref{eq:tpa} becomes singular at the points where $\dot{\psi}_j = -\dot{\psi}_{j - 1}$. Close to such a singularity, or when the $\psi,\theta$-grid is coarse, large or even non-monotonous time coordinates may result. When these events are identified,\footnote{``Large'' time steps are identified by FOX using the condition $|\dot{\psi}_j| > C|\psi_{j + 1} - \psi_j|/|t_{j + 1} - t_j|$, where the constant $C = 4$.} FOX successively eliminates points along the guiding center orbit until all time steps are small and monotonous. The $\phi_\rms$ coordinates of the orbit are then solved simply by using the trapezoidal method on eq.~\eqref{eq:tph}.

The eigenfunctions of the wave field, $\Phi_{i,m}(\psi)$ and the corresponding mode frequencies $\omega$ are solved by using the analytical approximations presented in Ref.~\cite{can}. Future versions of FOXTAIL intend to use the MHD eigenmode analyzer code MISHKA~\cite{mi1,mi2,mi3,mi4} to solve the TAEs numerically. Besides ideal MHD effects, MISHKA also considers effects from finite ion Larmor radii, ion and electron drift, neoclassical ion viscosity and bootstrap current, indirect energetic ion effects and the collisionless skin effect. The interaction coefficients $V_{i,\ell}$ are calculated from eq.~\eqref{eq:vil} on each point on the $\bm{J}$-grid also by using the trapezoidal method.

The user specifies which mode--Fourier coefficient pairs to be calculated by FOX. All bounce frequencies, precession frequencies, interaction coefficients, mode frequencies and toroidal mode numbers are then collected in a single output file. A separate output file is generated that contains all initial conditions used in a specific TAIL simulation, which contains the initial energetic particle distribution, flags on the mode--Fourier coefficient pairs to be active in the simulation, initial mode amplitudes, etc.

In the absence of collisions, the TAIL model equations of eq.~\eqref{eq:ta1} defines a system of ordinary differential equation, which is solved numerically using the standard 4$^\mr{th}$ order Runge-Kutta method. When momentum diffusion is present, the model equations instead become a system of stochastic differential equations, which can be modeled numerically, e.g.\ by using an It\=o--Taylor numerical scheme~\cite{klo}.

\section{Numerical studies}\label{sec:num}

\subsection{Comparison with the 1D bump-on-tail model}\label{sec:com}

One of the possible applications of FOXTAIL is to determine whether a one-dimensional bump-on-tail approximation of the system is sufficient to capture the essential wave--particle dynamics, such as growth rate and saturation amplitude. A higher computational efficiency typically follows from the lower dimensionality of the approximation. However, the 1D bump-on-tail model is only valid within certain parameter regimes. Section~\ref{sec:com} presents a quantitative numerical study of these regimes.

\subsubsection{Theoretical comparison}

A simple bump-on-tail model, neglecting collisions, explicit wave damping and particle sources and sinks, can be written on the form~\cite{tho}
\be\label{eq:bot}
	\dr{\xi_k}{\tau} = u_k,~\dr{u_k}{\tau} = \re(\tilde{A}\rme^{\rmi\xi_k}),~\dr{\tilde{A}}{\tau} = -\sum_k\rme^{-\rmi\xi_k},
\ee
where $\tau$ is the time coordinate, $\xi$ is the particle phase, $u$ is the particle momentum and $\tilde{A}$ is the complex eigenmode amplitude.

Now, consider the FOXTAIL model, only including a single mode $i$ and a single Fourier coefficient $\ell$. We define the particle phase as
\be
	\xi_k \equiv \ell\tilde{\theta}_k + n_i\tilde{\phi}_k - \omega_i t,
\ee
and the relative wave--particle frequency as
\be
	\Omega(\bm{J}_k) \equiv \dot{\xi}_k = \ell\omega_\rmB(\bm{J}_k) + n_i\omega_\rmp(\bm{J}_k) - \omega_i.
\ee
Next, we define a new momentum coordinate system $\bm{K} \equiv (\mu, S, W)$, where both $\mu$ and
\be
	S \equiv W - \frac{\omega_i}{n_i} P_\phi
\ee
are constants of motion as particles move along the wave--particle characteristic curves of mode $i$.

Assuming that the energetic particle distribution is located in a neighborhood around the wave--particle resonance $\bm{K}_\mr{res}$ (such that $\Omega(\bm{K}_\mr{res}) = 0$) where $\Omega' = \pt\Omega/\pt W$ and the interaction coefficient $V_{i,\ell}$ is constant in $\bm{K}$, the coordinate substitution
\be\label{eq:f2b}\br
	\tau \equiv t/\tilde{t}, & \tilde{A} \equiv \tilde{t}^2\Omega'(\bm{K}_\mr{res}) V_{i,\ell}(\bm{K}_\mr{res}) A_i, \\
	u_k \equiv \tilde{t}\Omega(\bm{K}_k), & \tilde{t} \equiv \left(\Omega'|V_{i,\ell}|^2\right)_{\bm{K}_\mr{res}}^{-1/3},
\er\ee
transforms the FOXTAIL model exactly to the 1D bump-on-tail model of eq.~\eqref{eq:bot}. Note that the substitutions of eq.~\eqref{eq:f2b} can be made for any FOXTAIL scenario, as long as one chooses a relevant eigenmode--Fourier coefficient pair and a resonant point $\bm{K}_\mr{res}$.

According to the theory of the 1D bump-on-tail model, the linear growth rate of the mode ($|\tilde{A}(\tau)| \approx \tilde{A}_0\rme^{\gamma_\rmL\tau}$) is~\cite{tho}
\be\label{eq:gam}
	\gamma_\rmL = \frac{\pi}{2}\left.\dr{\tilde{F}_0}{u}\right|_{u = 0},
\ee
where $\tilde{F}_0$ is the initial distribution function in $u$. The growth rate in time $t$ can be approximated by
\be\label{eq:gaf}
	\gamma_\rmL = \frac{\pi}{2}\frac{|V_{i,\ell}(\bm{K}_\mr{res})|^2}{\Omega'(\bm{K}_\mr{res})}\left.\dr{F_0}{W}\right|_{W_\mr{res}},
\ee
where $F_0$ is the $W$ distribution of energetic particles along the characteristic curves of wave--particle interaction ($F_0(W) = \int \rmd\mu\,\rmd S\,f_0(\bm{K})$). In the 1D bump-on-tail model, particles deeply trapped by the wave field oscillate in $\xi,u$-space around the resonance with the frequency\footnote{The frequency of eq.~\eqref{eq:omb} is in real time, $t$. For the frequency in the normalized time, $\tau$, the expression should be multiplied by $\tilde{t}$.}
\be\label{eq:omb}
	\omega_\rmb \approx \frac{\sqrt{|\tilde{A}|}}{\tilde{t}} = \sqrt{\left|\Omega'(\bm{K}_\mr{res})V_{i,\ell}(\bm{K}_\mr{res}) A_i\right|}.
\ee
By comparing the numerical growth rates and other system properties of the 1D bump-on-tail model and FOXTAIL, one can make estimations of the effects of having non-constant $\Omega'$ and interaction coefficients, and of having multiple modes and/or Fourier coefficients.

\subsubsection{Simulation parameters}\label{sec:sim}

In this study, we consider an ITER equilibrium configuration with an energetic particle distribution consisting of $^3$He$^{2+}$ ions. For simplicity, we neglect collisions, explicit wave damping and particle sources and sinks. In such scenarios, the amplitudes of the modes are expected to grow exponentially until saturation at an amplitude proportional to the growth rate squared \cite{lev}, assuming that the modes interact more or less independently with the energetic particle distribution. Since the magnetic moment is not a dynamical variable in this system, the dimensionality of the problem can be reduced by considering an energetic particle distribution on a $\mu = \mr{const.}$ surface. An ad hoc energetic particle distribution functions is constructed, with $\mu = \SI{0.5}{MeV/T}$ (\SI{1}{eV} =  \SI{1.6022e-19}{J}). The distribution in $W$ and $S$ are chosen to be Gaussians localized near the two resonances defined by $(n_i, \ell) = (5, 1)$ and $(n_i, \ell) = (6, 1)$.

\begin{figure}[t!]\centering
\includegraphics[width=88mm]{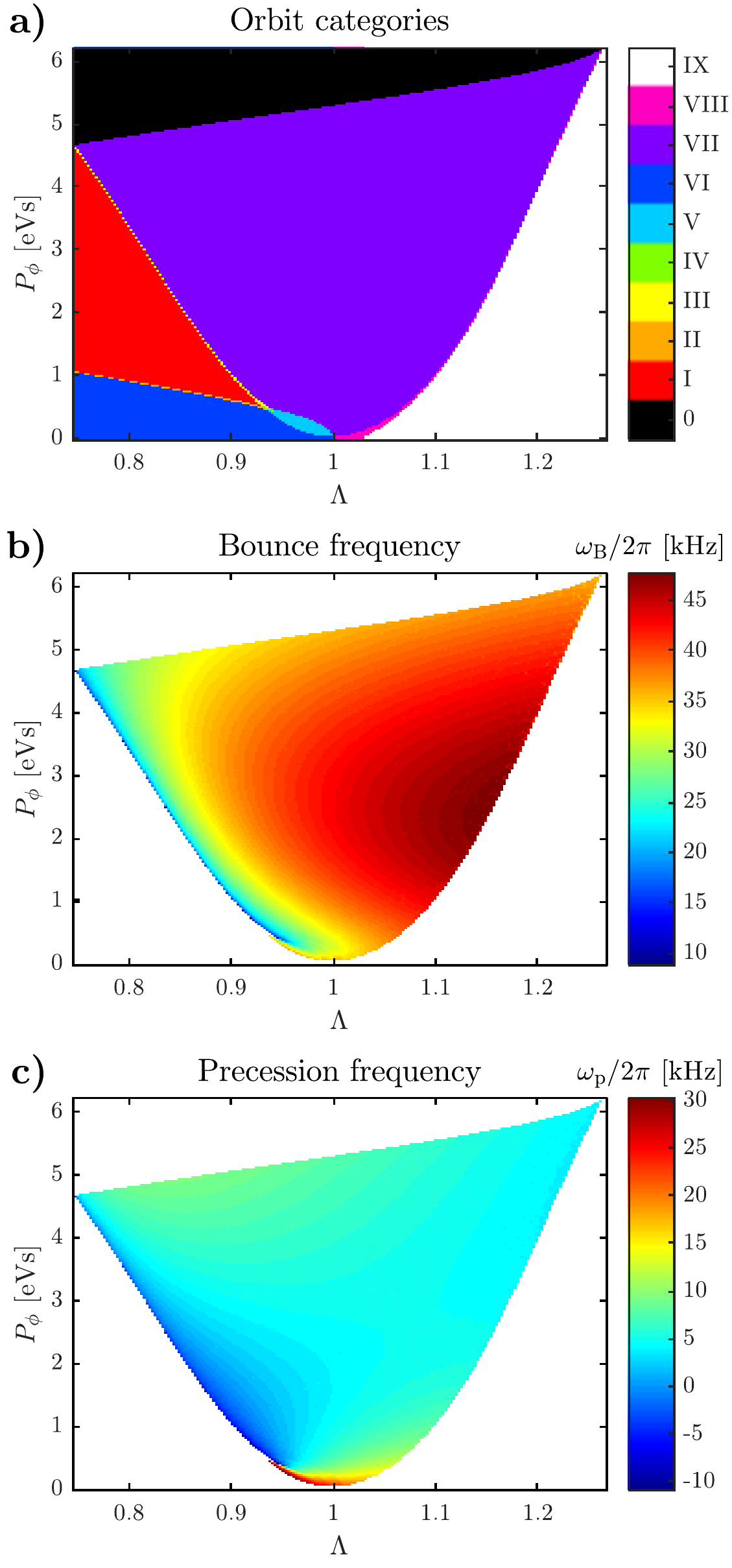}
\caption{The orbit categories, the bounce and the precession frequencies of trapped $^3$He$^{2+}$ ions on a $\Lambda,P_\phi$-grid with $\mu = \SI{0.5}{MeV/T}$. Category 0 orbits cross the last flux surface. For other categories, the convention of Ref.~\cite{hed} is used.}\label{fig:fbp}
\end{figure}

\begin{figure}[t!]\centering
\includegraphics[width=88mm]{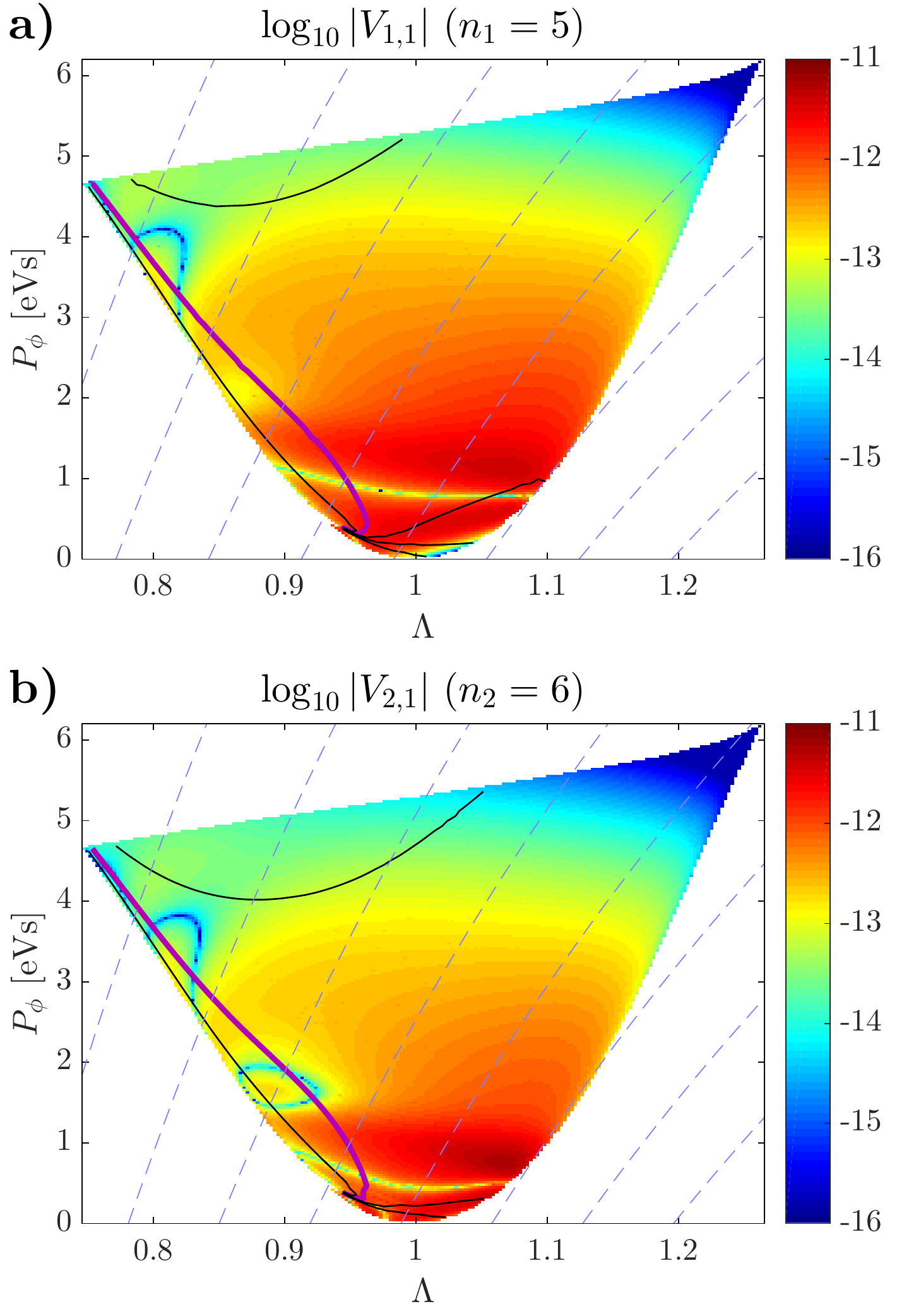}
\caption{The interaction coefficient $|V_{i,\ell}|$ plotted in $\bm{J}$-space, with $\mu = \SI{0.5}{MeV/T}$. The bold purple curves are the resonant curves $\ell\omega_\rmB + n_i\omega_\rmp = \omega_i$ for $\ell = 1$ and $i = 1$ (Fig.~\ref{fig:vil}.a), $i = 2$ (Fig.~\ref{fig:vil}.b). The thin black curves are the same resonant curves, but for $\ell = \{-2, -1, 0, 2\}$. The dashed curves are the wave--particle characteristic curves for the corresponding mode.}\label{fig:vil}
\end{figure}

\begin{figure}[t!]\centering
\includegraphics[width=88mm]{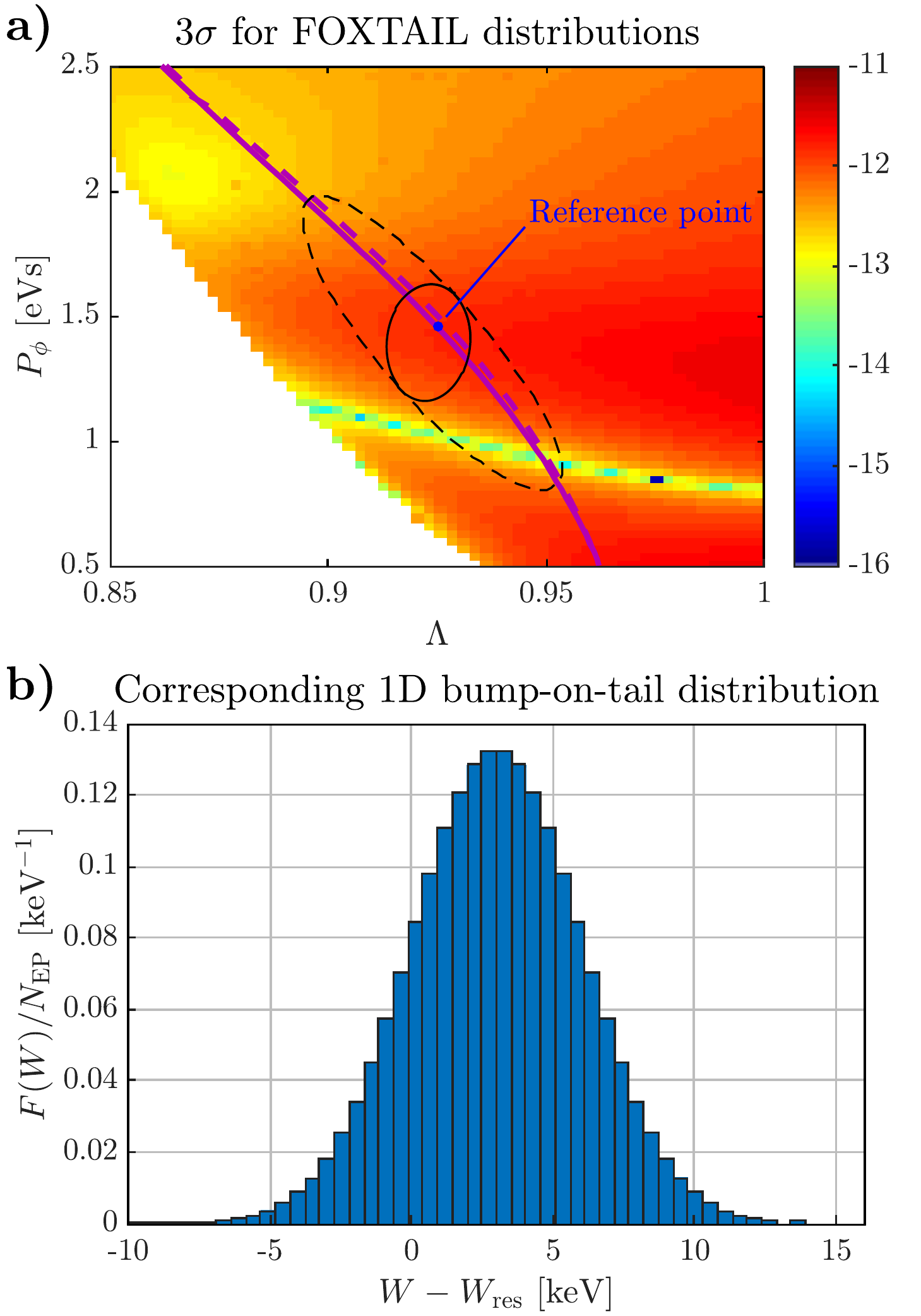}
\caption{The energetic particle distributions of this study is placed close to the $\ell = 1$ resonant curves. Figure~\ref{fig:dis}.a shows the 3$\sigma$ curves for the Gaussian distributions of the energetic particles (97\,\% of the particles are contained within the 3$\sigma$ curves). The solid black curve is the ``narrow'' initial distribution, and the dashed black curve is the ``wide'' initial distribution. The solid purple curve is the $(i, \ell) = (1, 1)$ resonance, and the dashed purple curve is the $(i, \ell) = (2, 1)$ resonance. The reference point (blue dot) used for transforming to the 1D bump-on-tail coordinate system is placed on the $(1, 1)$ resonance. Both initial distributions have the same corresponding 1D bump-on-tail distribution, shown in Fig.~\ref{fig:dis}.b, where $N_\mr{EP}$ is the total number of energetic particles. The Gaussian bump-on-tail distribution is centered at $W - W_\mr{res} = \SI{3}{keV}$, and it has a width of $\sigma = \SI{3}{keV}$.}\label{fig:dis}
\end{figure}

The chosen $\bm{J}$-grid to calculate the guiding center orbits is in the region of trapped particles (category V- and VII-orbits in Fig.~\ref{fig:fbp}.a) on the $\mu = \SI{0.5}{MeV/T}$ surface. The corresponding bounce and precession frequencies, presented in Figs.~\ref{fig:fbp}.b and \ref{fig:fbp}.c, respectively, are found from the time dependence of the guiding center orbits, which are solved using eqs.~\eqref{eq:tps} and \eqref{eq:tph}.

Two TAEs are chosen for this study: the first mode with a toroidal mode number $n = 5$, and the second one with $n = 6$. The energetic particle distribution is placed close to the resonance $(i, \ell) = (1, 1)$, i.e., the surface defined by $\omega_\rmB + 5\omega_\rmp = \omega_1$, where $\omega_1$ is the frequency of the first TAE. Figure~\ref{fig:vil} shows the calculated interaction coefficients $V_{i,\ell}(\bm{J})$ for $(i, \ell) = (1, 1)$, Fig.~\ref{fig:vil}.a, and $(i, \ell) = (2, 1)$, Fig.~\ref{fig:vil}.b. The frequencies of the two modes are \SI{38.2}{kHz} and \SI{40.2}{kHz}, respectively. Since the precession frequency is small compared to the bounce frequency at the $(1, 1)$ resonance, the $(1, 1)$ and the $(2, 1)$ resonances are almost the same in $\Lambda,P_\phi$-space.

A reference point is chosen on the $(1, 1)$ resonance, as shown in Fig.~\ref{fig:dis}.a. Around this point, the bounce and precession frequencies are approximated to first order in $W,P_\phi$-space, and $V_{1,1}$ is approximated to zeroth order, when transforming from the FOXTAIL to the 1D bump-on-tail coordinate system. Two different initial distribution functions are used in these studies: one localized around the reference point and one that is more spread along the resonance. The two distributions transform to the same distribution in the 1D bump-on-tail model, shown in Fig.~\ref{fig:dis}.b. The initial distributions are chosen such that there is a positive derivative of the energy distribution at the resonance, giving a positive linear growth rate according to eq.~\eqref{eq:gam}.

An energetic particle distribution consisting of $2.5\cdot 10^{16}$ $^3$He$^{2+}$ ions are distributed on $2.5\cdot 10^5$ markers. The markers are spread out in phase space using quasi-random low-discrepancy sequences (a Sobol' sequence~\cite{sob} combined with the Matou\v{s}ek scrambling method~\cite{mat}). They are placed as a Gaussian in $W$-space \emph{centered} around the resonance, and then the marker weights are set such that they represent the \emph{shifted} Gaussian as in Fig.~\ref{fig:dis}.b. This is done in order to improve the statistics of markers around the resonance as compared to a scenario with equal weights.

\begin{table*}\centering
%\hspace*{-23mm}
\begin{tabular}{cccccccc}
\toprule
Case & Eigenmodes & Fourier coefficients & $\sigma_S$ [keV] & $\sigma_W$ [keV] & $V_{2,1}$ factor & $\Delta\omega/2\pi$ [kHz] & $N_\mr{part}$ [$10^{16}$] \\
\midrule
\#1 & $i = \{1,2\}$ & $\ell = \{-2, \ldots, 2\}$ & 5.7 & 3.0 & 1 & 2.0 & 2.5 \\
\#2 & $i = 1$ & $\ell = 1$ & 5.7 & 3.0 & -- & -- & 2.5 \\
\#3 & $i = \{1,2\}$ & $\ell = \{-2, \ldots, 2\}$ & 22.7 & 3.0 & 1 & 2.0 & 2.5 \\
\#4 & $i = 1$ & $\ell = 1$ & 22.7 & 3.0 & -- & -- & 2.5 \\
\#5 & $i = \{1, 2\}$ & $\ell = 1$ & 5.7 & 3.0 & 10.1 & 2.0 & 2.5 \\
\#6 & $i = 2$ & $\ell = 1$ & 5.7 & 3.0 & 10.1 & -- & 2.5 \\
\#7 & $i = 1$ & $\ell = 1$ & 5.7 & 4.0 & -- & -- & 3.0 \\
\#8 & $i = \{1, 2\}$ & $\ell = 1$ & 5.7 & 4.0 & 13.2 & 6.1 & 3.0 \\
\#9 & $i = 2$ & $\ell = 1$ & 5.7 & 4.0 & 13.2 & -- & 3.0 \\\bottomrule
\end{tabular}\caption{Summary of the initial parameters used in the FOXTAIL simulations presented in Figs.~\ref{fig:bot} -- \ref{fig:na2}, where $\sigma_S$ and $\sigma_W$ are the energy widths of the Gaussian energetic particle distribution along the resonance curve and along the characteristic curves for wave--particle interaction, respectively, $\Delta\omega$ is the frequency separation between the two modes and $N_\mr{part}$ is the amount of energetic particles that the markers represent.}\label{tab:par}
\end{table*}

\subsubsection{Numerical comparison with the 1D bump-on-tail model}\label{sec:bot}

Here, different FOXTAIL scenarios are compared with the corresponding 1D bump-on-tail scenario by varying initial parameters such as the width of the initial energetic particle distribution along the resonances and the number of eigenmodes and Fourier coefficients to include in the simulations. Besides the 1D scenario, four FOXTAIL scenarios are presented. In the first two scenarios (referred to as cases \#1 and \#2), a narrow initial distribution is used (see Fig.~\ref{fig:dis}.a), whereas the two latter scenarios (cases \#3 and \#4) use the wide initial distribution. Furthermore, cases \#1 and \#3 include both of the TAEs presented in section~\ref{sec:sim} and a range of Fourier coefficients $-2 \leq \ell \leq 2$ for both eigenmodes. Cases \#2 and \#4 only include the first eigenmode ($n = 5$) and a single Fourier coefficient $\ell = 1$. For a complete list of initial parameter values used in the FOXTAIL simulations presented in this paper, see Table~\ref{tab:par}.

\begin{figure}[t!]\centering
\includegraphics[width=88mm]{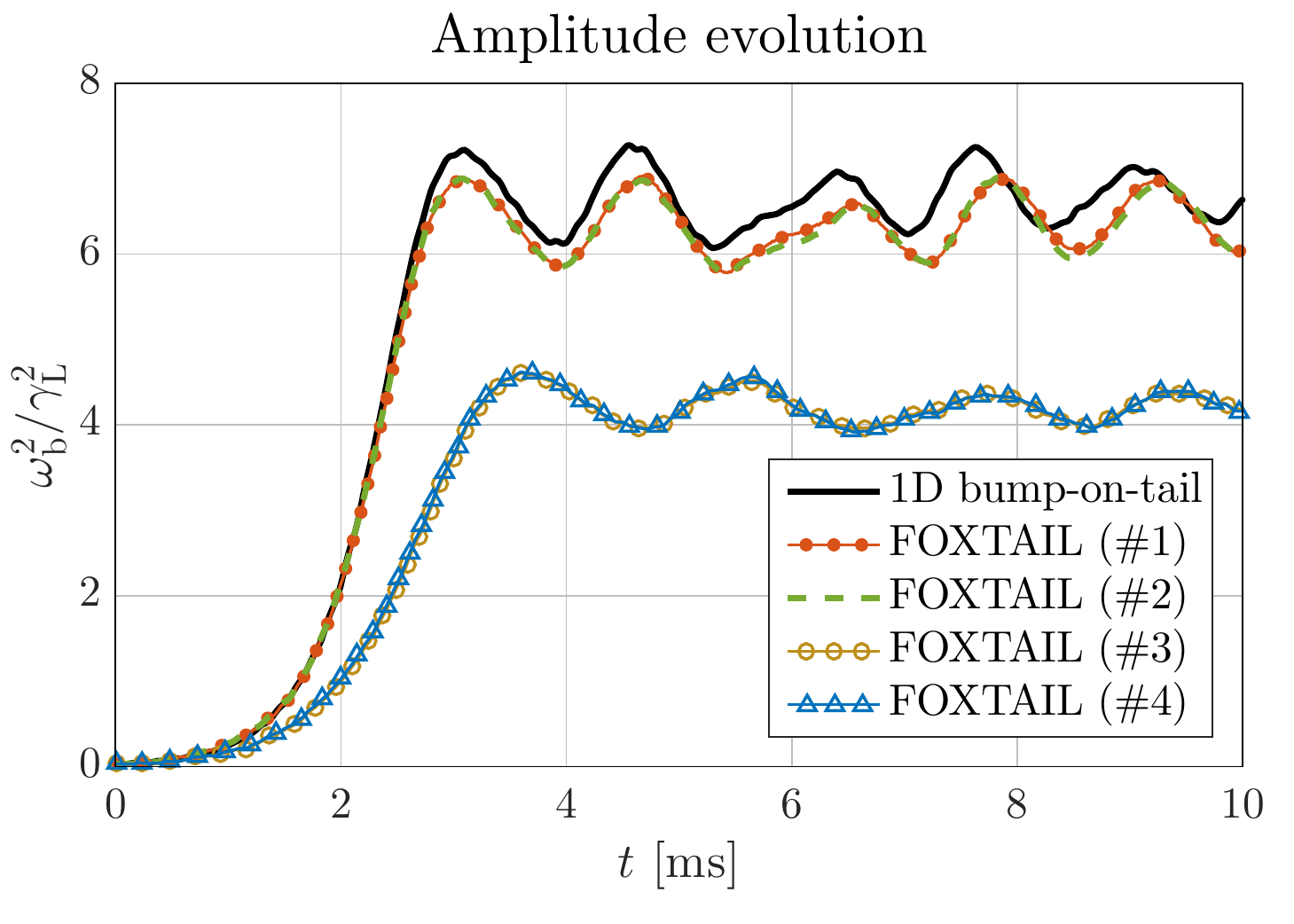}
\caption{Amplitude evolution of the first mode $(n = 5)$. $\omega_\mr{b} \propto \sqrt{|A_i|}$ is the bounce frequency of particles deeply trapped by the wave field, and $\gamma_\rmL$ is the analytical linear growth rate of the mode, calculated from the value of $|V_{i,\ell}|$ and $\pt\Omega/\pt W$ in the reference point. The solid black curve uses the 1D bump-on-tail model, which is analogous to FOXTAIL using a single mode--Fourier coefficient pair (in this case $(i, \ell) = (1, 1)$) and constant $V_{i,\ell}$ and $\Omega'$ in $\Lambda,P_\phi$-space. Cases \#1 and \#2 use a narrow initial energetic distribution function along the $(i, \ell) = (1, 1)$ resonance curve in $\Lambda,P_\phi$-space, whereas cases \#3 and \#4 use a wide initial distribution. Cases \#1 and \#3 include both TAEs and Fourier coefficients $-2 \leq \ell \leq 2$, whereas cases \#2 and \#4 only include the $(i, \ell) = (1, 1)$ mode--Fourier coefficient pair. See Table~\ref{tab:par} for a list of parameter values used in all FOXTAIL simulations.}\label{fig:bot}
\end{figure}

Figure~\ref{fig:bot} shows the amplitude evolutions of the first eigenmode for the bump-on-tail simulation and FOXTAIL simulations \#1 -- 4. The amplitude of the second mode never grows larger than $\approx$ 0.8\,\% of the amplitude of the first mode after saturation in case \#1, and up to 5\,\% in case \#3. It can immediately be seen that the two FOXTAIL scenarios with a narrow initial distribution (cases \#1 and \#2) agrees well with the corresponding 1D bump-on-tail scenario, both in growth rate and in saturation amplitude. On the other hand, the wider distribution (cases \#3 and \#4) gives a different growth rate and saturation level of the amplitude. This is presumably due to the fact that the wide distribution spans over regions where the interaction coefficient $V_{1,1}$ is considerably lower than in the reference point, giving a lower growth rate on average.

Including multiple eigenmodes and Fourier coefficients in the simulations seem to have negligible effects on the system in the considered scenarios, both for the narrow and the wide initial distributions. The reason for why the second mode does not influence the system considerably is because it has an expected growth rate of approximately 100 times lower than the first mode, which is due to the comparably lower values of the interaction coefficient $V_{2,1}$ in the region of the initial distribution function (recall that the growth rate scales as $|V_{i,\ell}|^2$, see eq.~\eqref{eq:gaf}).

When comparing growth rates of the different scenarios, it should be noted that $\gamma_\rmL$ of the 1D bump-on-tail scenario is approximately 81\,\% of the analytical growth rate of eq.~\eqref{eq:gam}, whereas the growth rates of cases \#1 and \#2 are 77\,\% of the analytical growth rate, and 68\,\% for cases \#3 and \#4. The reason why the growth rate of the 1D bump-on-tail scenario is considerably lower than the analytical one is primarily because of the finite extension of the energetic particle distribution along the characteristic curves (this issue was analyzed in more detail in Ref.~\cite{tho}).

\subsection{Numerical multi-mode studies}

\begin{figure}[t!]\centering
\includegraphics[width=88mm]{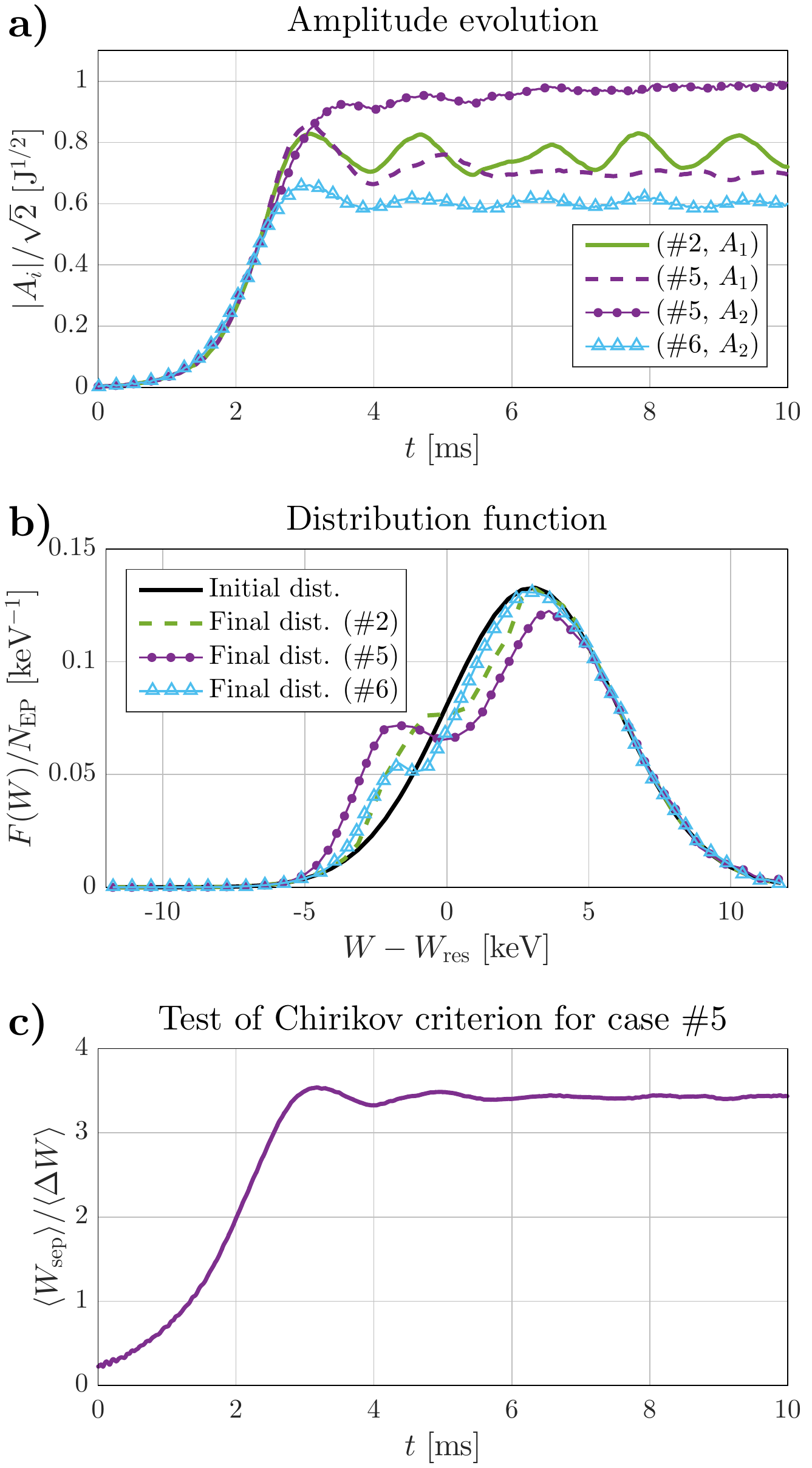}
\caption{Figure~\ref{fig:nar}.a shows the mode amplitude evolution for a set of FOXTAIL simulations using the narrow initial distribution function. Case \#4 includes both TAEs, and the Fourier coefficient $\ell = 1$ for each mode. Only the evolution of the first mode is presented. Case \#5 is the same as \#4, but the interaction coefficient $V_{2,1}$ is scaled up by a factor of 10.1, such that the linear growth rates of the two modes approximately match. Case \#6 is the same as \#5, but the first mode is deactivated in the simulation. See Table~\ref{tab:par} for a list of parameter values used in all FOXTAIL simulations. Figure~\ref{fig:nar}.b shows the corresponding 1D bump-on-tail distribution of the above simulations (the final distribution is at $t = \SI{10}{ms}$). Figure \ref{fig:nar}.c tests the Chirikov criterion for case \#5 by dividing the average resonance width in $W$-space and divide by the average energy separation between the resonances along the two characteristic curves.}\label{fig:nar}
\end{figure}

The presence of multiple eigenmodes proved to have negligible effect on the system in the scenarios presented in section~\ref{sec:bot} due to the low interaction coefficient of the second mode in the considered part of $\Lambda,P_\phi$-space, compared to the interaction coefficient of the first mode. Scenarios with significant multimode dynamics can be constructed by adding an ad hoc scaling factor to the interaction coefficient of the second mode. Multiplying $V_{2,1}$ by a factor of 10.1 gives approximately the same linear growth rate of the two modes. This has been done for cases \#5 and \#6, presented in Fig.~\ref{fig:nar}, along with the previous case \#2. The three scenarios are the same, except that in cases \#2 and \#6 the eigenmodes are simulated individually, whereas in case \#5 both modes are included to the system. Such a comparison allows one to isolate the nonlinear effects of mode interaction via the energetic particle distribution.

\begin{figure}[t!]\centering
\includegraphics[width=88mm]{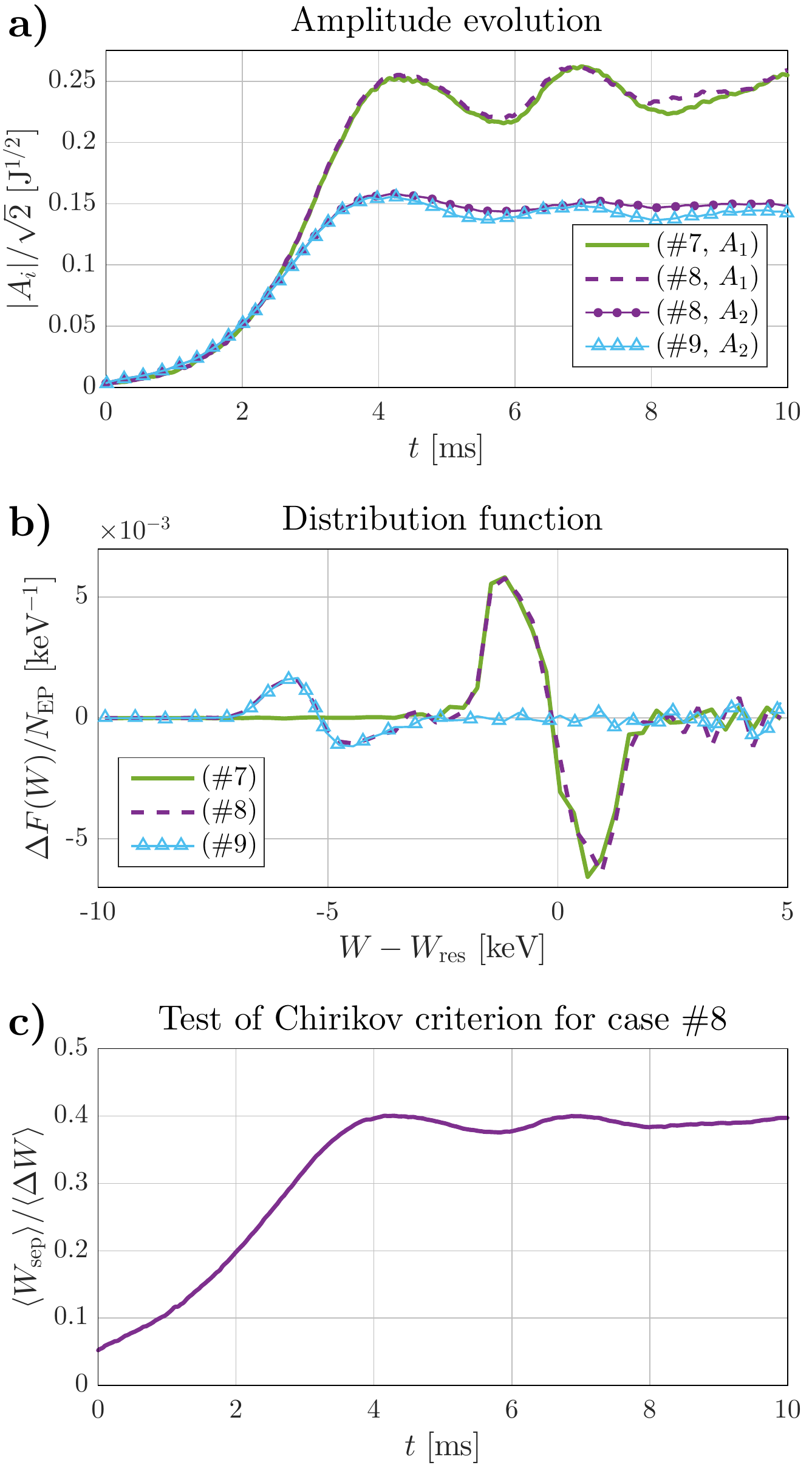}
\caption{The same as Fig.~\ref{fig:nar}, but for slightly different scenarios. The frequency separation between the two modes is increased by a factor of 3, the width of the Gaussian energetic particle distribution along the characteristic curve of the first mode is increased from $\sigma = \SI{3}{keV}$ to $\sigma = \SI{4}{kev}$ and the number of particles is increased from $2.5\times 10^{16}$ to $3.0\times 10^{16}$. The scaling factor of the interaction coefficient $V_{2,1}$ is increased from 10.1 to 13.2 in order to match the linear growth rates of the two modes. The amplitude evolutions of the second mode ($A_2$) in Fig.~\ref{fig:na2}.a are smoothened in order to remove high frequency amplitude oscillations coming from the interactions with off-resonant particles and from statistical fluctuations. $\Delta F(W)$ of Fig.~\ref{fig:na2}.b is the difference between the final and the initial corresponding bump-on-tail distributions.}\label{fig:na2}
\end{figure}

Comparing the multimode scenario with the scenarios where the modes are included individually, it can be seen that the indirect interaction between the modes via the energetic particles has significant macroscopic effects on the system. This can partly be understood as a consequence of stochastization of particle trajectories in phase space due to resonance overlap of the two eigenmodes. Stochastization of trajectories causes a locally enhanced transport of energetic particles around the resonances, allowing the eigenmodes to exhaust more energy from the energetic particle distribution (see e.g.\ stochastization from resonance overlap, \cite{bb4,bb5}, and from phase decorrelation, \cite{tho}). This results in a wider portion of the energetic particle distribution being flattened around the resonances, compared to the individual eigenmode cases, as seen in Fig.~\ref{fig:nar}.b.

The resonance width of an eigenmode can be estimated as the separatrix width of the unperturbed mode (i.e.\ in the absence of other modes) along the characteristic curve in $W$-space, using the 1D bump-on-tail approximation. When the resonance widths of the two modes are comparable, the resonance-overlap parameter~\cite{chi} (English translation: Ref.~\cite{ch2}), estimated as the average resonance width of the two modes divided by their distance in phase space, is an approximate measure of the level of stochastization of particle trajectories. The Chirikov criterion for stochastization is satisfied when the resonance-overlap parameter is larger than unity. The full separatrix width in $W$-space, $W_\mr{sep}$, is $4\omega_\rmb/\Omega'(\bm{K}_\mr{res})$, with $\omega_\rmb$ given by eq.~\eqref{eq:omb}. As seen in Fig.~\ref{fig:nar}.c, the Chirikov criterion is well satisfied for case \#5 after $t \gtrsim \SI{1.3}{ms}$.

Slightly different scenarios are tested in the simulations presented in Fig.~\ref{fig:na2}. Cases \#7, \#8 and \#9 are the same as cases \#2, \#5 and \#6, respectively, except that the frequency separation between the two eigenmodes is artificially increased by a factor of 3, and the width of the energetic distribution function along the characteristic curves of the first mode is increased to $\sigma = \SI{4}{keV}$ instead of \SI{3}{keV}. The scaling factor of the interaction coefficient is also adjusted such that the linear growth rates of the two modes approximately match. As can be seen in Fig.~\ref{fig:na2}.c, the Chirikov criterion is never satisfied for case \#8, although the resonance-overlap parameter is of the order of unity. Comparing the amplitude evolutions in Figs.~\ref{fig:nar}.a and \ref{fig:na2}.a, the two modes of case \#8 evolves more similarly to the corresponding individual mode scenarios than case \#5 does. This is especially seen in Fig.~\ref{fig:na2}.b, where the modes of case \#8 flattens two separate regions of the energetic distribution function, matching the flattening regions of the individual mode scenarios. 

\section{Summary}\label{sec:con}

This paper presents the theoretical framework of the FOXTAIL code, which is used to describe the nonlinear interaction between Alfvén eigenmodes and energetic particles in toroidal geometries. FOXTAIL is a hybrid magnetohydrodynamic--kinetic model based on a model developed by Berk \etal~\cite{bb1}, where each simulation is formulated as an initial value problem. Eigenmodes are treated as perturbations of the equilibrium system, with temporally constant eigenfunctions and dynamic complex amplitudes that vary on time scales longer than the inverse mode frequency. The energetic particle distribution is modeled by a finite set of markers in an action--angle phase space of the unperturbed system. The use of action--angle coordinates rather than conventional toroidal coordinates simplifies the equations of motion of the individual markers, and it allows for efficient resolution of time scales relevant for resonant eigenmode--particle interaction in numerical simulations.

The particle response with respect to the wave field is quantized by a Fourier series expansion of the kinetic energy change $q \bm{v}\cdot\delta\bm{E}$ along the transit period of the unperturbed guiding center orbit, where $q$ is the particle charge, $\bm{v}$ is the guiding center velocity and $\delta\bm{E}$ is the electric wave field at the guiding center position. A Lagrangian formulation of the wave--particle system, consistent with the derived particle response with respect to the wave, is used to derive equations for the eigenmode amplitudes and phases. The resulting system of equations describing direct wave--particle interaction has a phase space with four particle dimensions and two eigenmode dimensions (amplitude and phase). When including mechanisms that perturb the magnetic moment of energetic particles, the particle phase space extends to five dimensions.

When splitting the interaction in the Fourier terms along the transit period, each term contributes to resonant nonlinear interaction mainly in a narrow region around surfaces in the adiabatic invariant space, referred to as resonant surfaces. These surfaces are given by $\ell\omega_\rmB + n_\phi\omega_\rmp = \omega_\mr{mode}$, where $\ell$ is the Fourier index of interaction, $\omega_\rmB$ is the bounce frequency, $n_\phi$ is the toroidal mode number of the eigenmode, $\omega_\rmp$ is the precession frequency and $\omega_\mr{mode}$ is the eigenmode frequency. The width of the relevant region around the resonant surfaces depends on the amplitude of the eigenmode, the strength of wave--particle interaction at the resonant surfaces (quantified by the Fourier coefficients of interaction) and the variation of bounce and precession frequencies of particles along the characteristic curves of wave--particle interaction (the curves in adiabatic invariant space along which a given eigenmode accelerates particles).

The presented multi-dimensional model can be approximated with a conventional 1D bump-on-tail model. For the 1D approximation to be valid, three approximate criteria must be met:
\begin{itemize}
\item No more than one eigenmode--Fourier coefficient pair interacts significantly with the energetic particle distribution.
\item The complex Fourier coefficient of interaction is approximately constant in adiabatic invariant space throughout the resonant part of the energetic particle distribution.
\item The bounce and precession frequencies of the energetic particles vary approximately linearly in kinetic energy--toroidal canonical momentum space across the region of the resonance where the energetic particle distribution is located.
\end{itemize}
All these conditions can be quantitatively evaluated with FOXTAIL.

Effects of the fulfillment of the Chirikov criterion in scenarios with two active eigenmodes have been studied numerically using FOXTAIL. It has been verified that eigenmodes can be treated independently in scenarios where the criterion is not satisfied. On the other hand, when the resonance-overlap parameter becomes larger than unity, indirect mode--mode interaction via the energetic particle distribution becomes significant, and a larger portion of the inverted energetic particle distribution becomes flattened in energy space due to stochastization of particle trajectories in phase space.



\section*{Acknowledgments}
This work was supported by the Swedish research council (VR) contract 621-2011-5387.

%\section*{References}
\bibliographystyle{elsarticle-num}
\bibliography{CPC_ref}

\end{document}